\documentclass[letterpaper,twocolumn,prl]{revtex4-1}
\usepackage{mathptmx}
\usepackage[latin9]{inputenc}
\usepackage{color}
\usepackage{amsmath}
\usepackage{amssymb}
\usepackage{graphicx}
\usepackage{esint}
\usepackage[unicode=true,pdfusetitle,
 bookmarks=true,bookmarksnumbered=false,bookmarksopen=false,
 breaklinks=false,pdfborder={0 0 0},pdfborderstyle={},backref=false,colorlinks=true]
 {hyperref}
\hypersetup{
 colorlinks,linkcolor=red,citecolor=blue}
\usepackage{breakurl}

\makeatletter

\usepackage{xcolor}\usepackage{soul}

\definecolor{blue}{rgb}{0,0,1}
\definecolor{red}{rgb}{1,0,0}
\definecolor{green}{rgb}{0,1,0}

\makeatother

\begin{document}
\title{Infrared Laser Locking to Rubidium Saturated Absorption Spectrum via a Photonic Chip Frequency Doubler}

\author{Jiacheng Xie$^{1,2}$, Jia-Qi Wang$^{1,2}$, Zhu-Bo Wang$^{1,2}$, Xin-Xin Hu$^{1,2}$,Xiang Guo$^{3}$, Rui Niu$^{1,2}$, Joshua B. Surya$^{3}$, Ji-Zhe Zhang$^{1,2}$, Chun-Hua Dong$^{1,2}$, Guang-Can Guo$^{1,2}$, Hong X. Tang$^{3}$, Chang-Ling Zou$^{1,2,*}$}

\affiliation{$^{1}$Key Laboratory of Quantum Information, University of Science and Technology of China, Hefei, 230026, People's Republic of China}
\affiliation{$^{2}$Synergetic Innovation Center of Quantum Information \& Quantum Physics, University of Science and Technology of China, Hefei, Anhui 230026, China}
\affiliation{$^{3}$Department of Electrical Engineering, Yale University, New Haven, Connecticut 06511, USA}
\affiliation{$^{*}$Corresponding author: clzou321@ustc.edu.cn}

\begin{abstract}
Photonic integrated resonators stand out as reliable frequency converters due to their compactness and stability, with second-harmonic generation (SHG) efficiencies of up to 17000$\%/W$ reported recently in aluminum nitride microrings. In this work, a sufficiently strong second-harmonic (SH) signal up to microwatts was generated by a photonic integrated frequency doubler using a milliwatt infrared (IR) laser source. Furthermore, increased SHG bandwidth covering $^{85}$Rb and $^{87}$Rb D$_2$ transition lines as well as saturated absorption spectroscopy (SAS) were demonstrated by tuning the pump power and chip temperature. Here, we present, to the best of our knowledge, the first successful locking of an IR laser to Rb saturated absorption lines via a photonic chip frequency doubler.
\end{abstract}

\date{\today}
\maketitle

\begin{figure*}[htbp]
\centering
\includegraphics[trim={0 3cm 0cm 0cm},clip,width=0.7\linewidth]{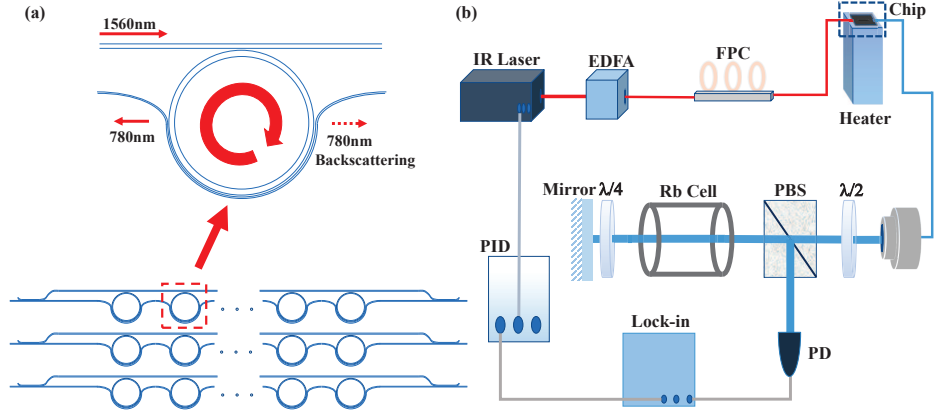}
\caption{(a) Schematic illustration of AlN microring resonators, where eight microrings share the same set of bus waveguides, but have slightly different design parameters. Ideally, SH light is coupled out through the lower wrap-around waveguide and transmits in the opposite direction of pump light. Backscattered SH light can also be observed in the reverse direction. (b) The experimental setup for laser locking. The chip is placed on top of a heater. Pump light from an IR laser is coupled into one side of the chip using a lensed fiber, and SH light is coupled out from the other side. This SH beam will further be collimated into free space for SAS and laser locking.}
\label{1}
\end{figure*}

Quantum transitions are widely used as optical frequency references (OFRs) in modern physics \cite{0953-4075-38-9-003}, since they meet a crucial requirement for frequency standards, i.e., reproducibility \cite{timefrequency}, which originates from the stable energy difference between electronic quantum states. Since the invention of lasers, the utilization of quantum transitions in laser-based experiments for the purpose of frequency stabilization has become highly ubiquitous \cite{Arie:92,doi:10.1063/1.1654607,Lucero:91,Breton,yan,sakai,deLabachelerie:95}. For lasers in the low-loss transmission window of optical fibers around 1550$\,$nm, achieving a fixed frequency has always been a topic of interest. Both atomic \cite{Lucero:91,Breton} and molecular \cite{yan,sakai,deLabachelerie:95} transitions were proposed as potential OFRs. However, the weak overtone or combination bands of molecular transitions \cite{Mahal:96} and the additional excitations \cite{Chung,Lucero:91,Breton} of atomic transitions in the corresponding wavelength ranges add to the difficulties for these OFRs to achieve Doppler-free SAS \cite{1996AmJPh..64.1432P}.

One way to circumvent this problem is by using a frequency doubler to up-convert the laser frequency to visible frequency, which extends the coherence from visible wavelengths to telecom wavelengths. For example, we could lock an IR laser to 1560.48$\,$nm by locking its SH light to the Rb D$_2$ saturated absorption lines. Recent works have explored related ideas \cite{Mahal:96,Zhu:97,Bruner:98}. However, the free-space systems used were generally lacking in compactness. Thus, we propose the use of a photonic integrated circuit (PIC) as a highly compact and stable platform for frequency doubling, where the high quality (Q) factor resonances of microring resonators would greatly enhance the SHG conversion efficiency. Recently, SHG efficiencies up to 2500\%/W \cite{Guo:16} and 17000\%/W \cite{Alex} were reported on PICs, respectively with polycrystalline and crystalline aluminum nitride microring resonators. Such high conversion efficiencies enable us to achieve a microwatt SH output with milliwatt input, which lays the foundation for Rb SAS \cite{Zhu:97,643293}.

For the realization of SHG in microring resonators, the frequency of the fundamental mode $\omega_a$ should be about half the frequency of the SH mode $\omega_b$. The efficiency reaches its maximum when $\omega_b=2\omega_a$ under ideal phase matching conditions \cite{Guo:16}. Thus, there exists a challenge for practical applications of such an integrated frequency doubler for laser locking, i.e., the SH mode of the integrated chip should be exactly on-resonance with the atomic transition, and simultaneously the half frequency of the Rb transition should match the fundamental mode. Another challenge is that normally the bandwidth of such resonators is very small, which limits the reliability of the frequency-doubler-based laser locking, since the laser frequency may drift out of the resonance and break the feedback control loop. Therefore, to achieve the goal of locking an IR laser to Rb SAS via an integrated frequency doubler, we have to solve the two doubly-resonant problems described above.

In this letter, we demonstrate successful locking of an IR laser to Rb saturated absorption lines, using the SH signal from an integrated microring resonator. By tuning the temperature and pump power of the chip, we are able to achieve doubly-resonant enhanced SHG precisely at the Rb D$_2$ absorption lines. We also propose to utilize the thermal bistability effect of the microring resonator for extending the SHG bandwidth, which allows us to reliably lock an IR laser to one of the Rb saturated absorption lines. We anticipate that the demonstrated compact SHG-based IR laser locking system would find applications in various integrated atomic clock systems \cite{newman2018photonic}, while the thermal mechanism demonstrated in this work could also be generalized to other integrated nonlinear optical system \cite{luo2018semi,wolf2018quasi,wang2018ultrahigh,logan2018400,chang2018heterogeneously}.

Figure$\,$\ref{1}(a) is a schematic illustration of the integrated microring frequency doubler. Eight microrings are connected using one set of bus waveguides, each with different radius and ring width. The top point-contact bus waveguide is used to transmit IR pump light, while the lower wrap-around waveguide is used to transmit visible SH light. On-chip wavelength division multiplexers (WDM) at both ends are used to couple the set of bus waveguides. The details about the design and fabrication process can be found in Ref.$\,$\cite{xiangpra}. When the doubly-resonant conditions are met for SHG, the IR pump light sent to the top point-contact bus waveguide will be converted to SH light and collected by the lower wrap-around waveguide. To simplify our experimental setup, we collected the backscattered SH light propagating along the same direction as the transmitted pump light. 

The experimental setup for locking an IR laser to Rb saturated absorption lines is illustrated in Fig.$\,$\ref{1}(b). An IR laser was amplified by an erbium-doped fiber amplifier (EDFA) and coupled to the chip, while the polarization of the input light was optimized by a fiber polarization controller (FPC) in order to achieve the highest SHG efficiency at a specific temperature $T$. Once the doubly-resonant conditions for SHG were satisfied, SH light was generated and collected by a fiber at the output port. To stabilize the IR laser, the SH light was compared to the Rb atomic transitions generated by a SAS setup, which consisted of a half-wave plate, a polarizing beamsplitter (PBS), a Rb cell, a quarter-wave plate and a mirror. The SAS was detected by a silicon photodetector (PD). By virtue of a proportional-integral-derivative (PID) controller and a lock-in amplifier, we were able to lock the laser wavelength to a specific peak in the SAS.

\begin{figure}[t]
\centering
\includegraphics[trim={0cm 0cm 0cm 0cm},clip,width=\linewidth]{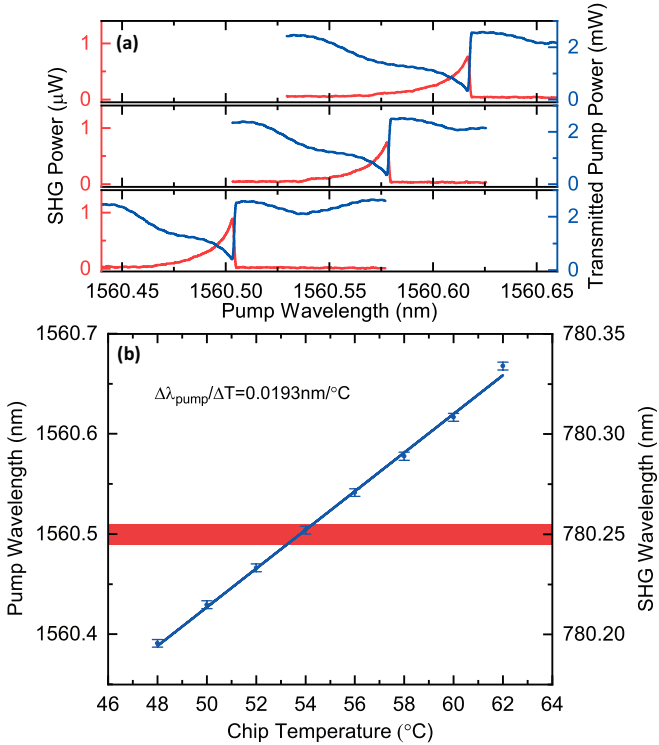}
\caption{Temperature dependence of SH signals. (a) Red traces represent SHG power and blue traces represent transmitted pump power. The chip temperatures for the three subfigures (from top to bottom) are 60$^\circ$C, 58$^\circ$C, 54$^\circ$C respectively. Off-chip pump input is set to about 10.9$\,$mW. (b) Temperature dependence of resonant wavelength (at which the SH signal reaches its highest intensity). Off-chip pump input is set to about 10.9$\,$mW. The red region covers SH wavelengths in the range [780.245, 780.255]$\,$nm. The corresponding temperature range is [53.2, 54.3]$\,^\circ$C. The slope reads 0.0193$\,$nm/$^\circ$C.}
\label{fig2}
\end{figure}

The IR transmission and SH signals were collected by scanning the input IR laser wavelength. Due to the abundance of microrings on the chip, one with SHG phase-matching wavelength closest to the Rb D$_2$ transitions was selected, with typical results plotted in Fig.$\,$\ref{fig2}(a). We noticed that the spectral shapes of the IR transmissions and SH signals were significantly asymmetric under $10.9\,$mW off-chip pump input, which was attributed to the thermal bistability effect in a high Q microcavity \cite{Carmon:04}. As the laser approached the resonant wavelength, the extinction of the IR input increased, and the temperature of the microring increased as well due to a higher intracavity laser power. Simultaneously, the power of the SH light also increased due to a higher absorption of pump light. By varying the chip temperature, we observed a wavelength shift of the SH signal, since both fundamental and SH modes of the microring were shifted. As described by Fig.$\,$\ref{fig2}(a), it was observed that as the temperature increased, the SH signals shifted to longer wavelengths while the conversion efficiency decreased. The shift of the resonant wavelength can be explained by the thermo-optic effect and thermal expansion \cite{Carmon:04}, whereas the temperature dependence of SHG efficiency results from the differing thermo-shift coefficients of the fundamental and SH modes, since the conversion efficiency is dependent on the mismatch between the fundamental and SH mode resonant wavelengths. Such temperature dependence of SHG wavelength offers a convenient way to match the resonant frequency doubler with a specific wavelength. The relationship between the resonant wavelength and the chip temperature with a fixed IR laser power is summarized in Fig.$\,$\ref{fig2}(b), where a linear relationship between SHG wavelength and chip temperature is demonstrated. Wavelengths that lie in the red region are near the Rb absorption lines. The corresponding temperature range gives us practical values for conducting the laser-locking experiment.

\begin{figure}[t]\centering
\includegraphics[trim={0cm 0cm 0cm 0cm},clip,width=0.94\linewidth]{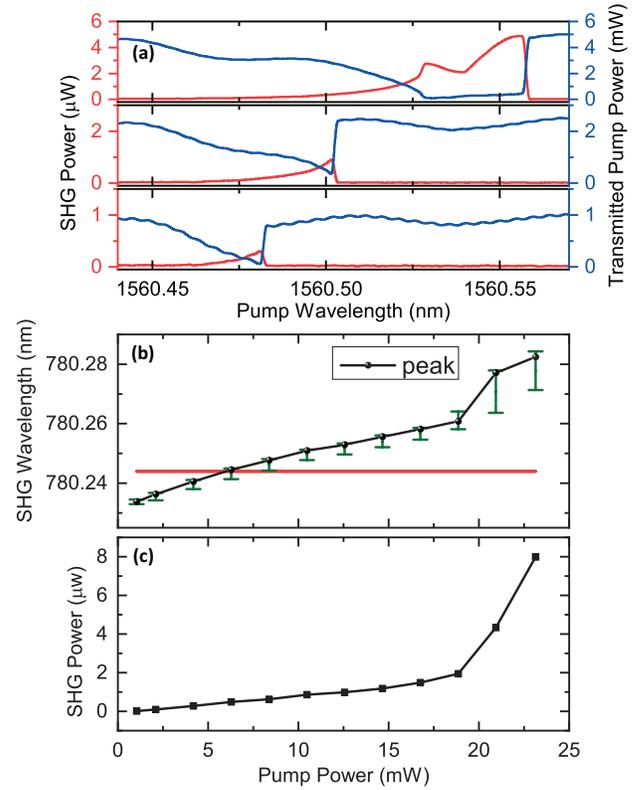}
\caption{Pump dependence of SH signals. (a) Red traces represent SHG power and blue traces represent transmitted pump power. The off-chip pump powers for the three subfigures (from top to bottom) are 21.0$\,$mW, 10.5$\,$mW, 4.2$\,$mW respectively. The chip temperature is 54$^\circ$C. (b) Pump dependence of SHG wavelength. The dot represents the wavelength at which the SHG power reaches its peak. The vertical bar gives the wavelength range in which the SHG power exceeds half of the peak power. (c) Pump dependence of SHG power. The dot gives the peak power of output SH light for a specific off-chip pump power.}
\label{fig3}
\end{figure}

For our devices, the typical linewidth of the IR mode is less than $0.5\,$GHz. It seems that such a narrow linewidth of on-chip microring would impose a limit of SHG bandwidth, because it is comparable with the Doppler-broadened absorption linewidth of Rb. However, it was observed that the thermal bistability effect would induce an expansion of the SHG bandwidth, as indicated in Fig.$\,$\ref{fig2}(a). Therefore, the thermal bistability effect can assist a full coverage of $^{85}$Rb and $^{87}$Rb D$_2$ absorption lines, and makes it possible to reliably lock the IR laser to SAS and prevents laser from drifting out of the resonance. Fig.$\,$\ref{fig3} summarizes the power dependence of the SH signals from the microring. Similarly to Fig.$\,$\ref{fig2}(a) where an increase in temperature shifts the SH signal to a longer wavelength, in Fig.$\,$\ref{fig3}(a), we notice that as the pump power increases, SH signals also shift in the same direction. This is a well-known thermal effect due to the large pump power, i.e., a higher pump power will heat the microring cavity to a higher temperature. Fig.$\,$\ref{fig3}(b) depicts the relationship between SHG wavelength (peak and full width at half maximum (FWHM)) and pump power, with the red line indicating $^{85}$Rb D$_2$ F=3 absorption lines (780.244$\,$nm). The intersection between the red line and the experimental curve gives optimal pump power for the laser-locking experiment for a chip temperature of 54$^\circ$C. The vertical bar of Fig.$\,$\ref{fig3}(b) illustrates the SHG bandwidth corresponding to the FWHM of SH signals, which shows the trend that the SHG bandwidth increases with pump power. Fig.$\,$\ref{fig3}(c) depicts the relationship between SHG power and pump power. Normally, the SHG efficiency $P_{\rm{SHG}}/P_p$ saturates at high pump power regions \cite{Guo:16}. However, as shown in Fig.$\,$\ref{fig3}(c), rather than showing a saturation behavior, the curve suddenly undergoes a drastic increase when the pump power reaches around $20\,$mW. The FWHM (represented by vertical bar) and the peak wavelength (Fig.$\,$\ref{fig3}(b)) show similarly sudden changes when increasing the pump power. This phenomenon is due to the doubly-resonant condition being satisfied for SHG at a certain temperature, which can be determined by both the chip temperature and the pump power.

\begin{figure}[t]\centering
\includegraphics[trim={0.5cm 0cm 3cm 0cm},clip,width=\linewidth]{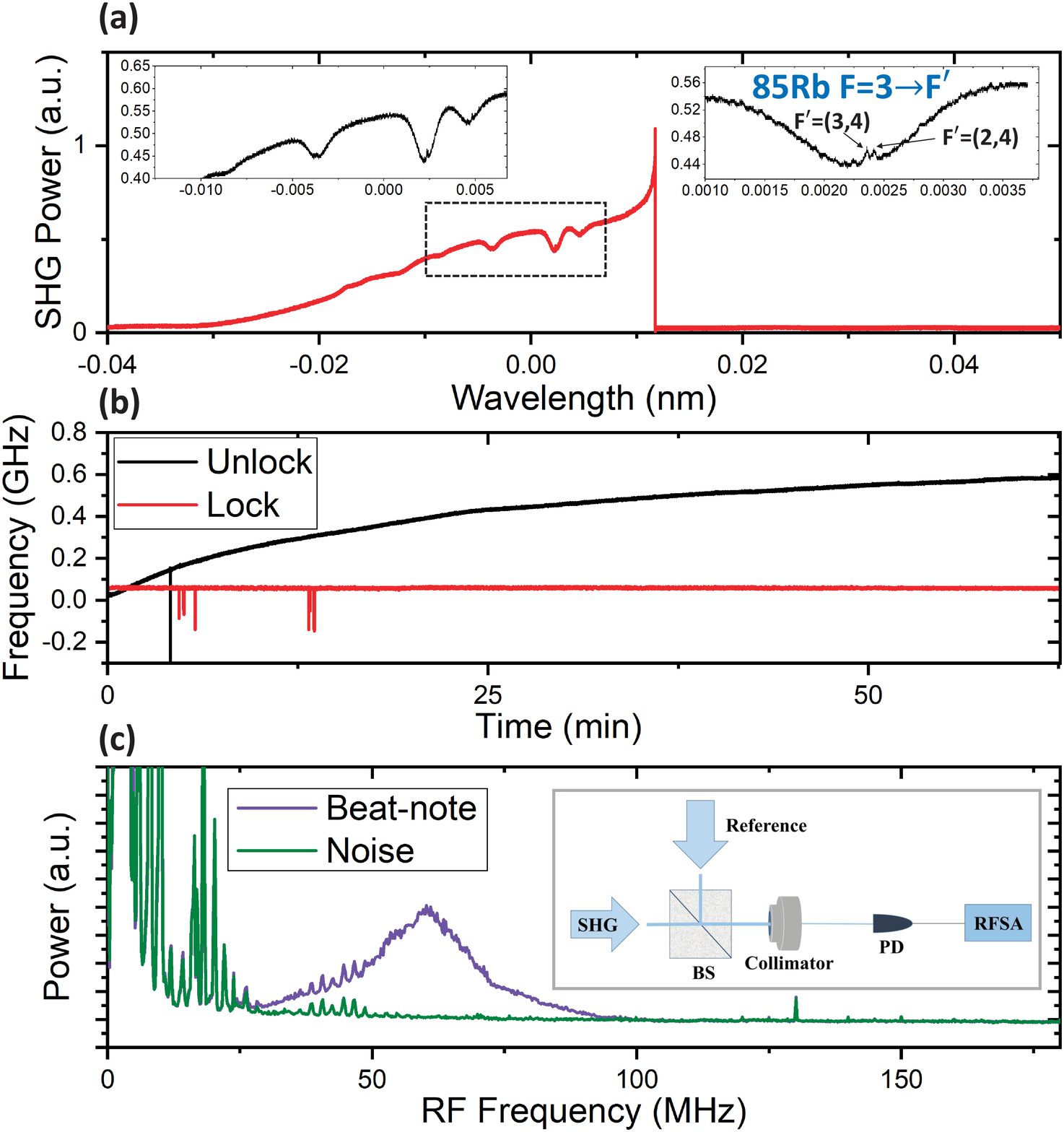}
\caption{(a) Rb saturated absorption of SH signal. Inset on the right: zoom-in of saturated absorption lines $^{85}$Rb F=3$\rightarrow$F'=(3,4) and F=3$\rightarrow$F'=(2,4). They belong to $^{85}$Rb D$_2$ transitions. (b) The black trace represents the drift of laser frequency without locking and the red trace represents the stablized laser frequency after locking the SH signal to a saturated absorption line. The sudden change of laser frequency is due to the readout error of wavemeter. (c) Beat-note frequency spectrum and setup schematic illustration. The reference light from a visible laser is locked to the saturated absorption line $^{85}$Rb F=3$\rightarrow$F'=4. The SH light is locked to the crossover line $^{85}$Rb F=3$\rightarrow$F'=(3,4). These two lines have a frequency difference of 60.3$\,$MHz.}
\label{fig4}
\end{figure}

Based on these studies, we were able to achieve a wide-bandwidth, high-intensity SH output that covered $^{85}$Rb and $^{87}$Rb D$_2$ absorption lines by tuning both pump power and chip temperature. Fig.$\,$\ref{fig4}(a) shows the Rb SAS by using SH signals from our integrated frequency doubler, with an off-chip pump power of around 80mW, and a temperature of 48$^\circ$C. The zoom-in figure shows the saturated absorption lines of $^{85}$Rb F=3$\rightarrow$F'=(3,4) and F=3$\rightarrow$F'=(2,4), where the bracket denotes crossover lines. An IR laser was successfully locked to these saturated absorption lines using a dither locking technique. The uncertainty of the IR laser after locking the SH signal to the SAS was verified to be about the wavemeter resolution of 20$\,$MHz, a large improvement in comparison with the drift of 600$\,$MHz for an unlocked laser as shown in Fig.$\,$\ref{fig4}(b). For further confirmation, we used an external-cavity diode visible laser which was locked to the saturated absorption line $^{85}$Rb F=3$\rightarrow$F'=4 as the reference, while the SH signal from the microring was locked to the saturated absorption line $^{85}$Rb F=3$\rightarrow$F'=(3,4), and measured the beating between the SH light of the locked IR laser and the reference laser. Shown in the inset of Fig.$\,$\ref{fig4}(c) is the experimental setup, alongside the beating signal spectrum detected by a radio-frequency spectrum analyzer (RFSA). The beat-note frequency spectrum in Fig.$\,$\ref{fig4}(c) shows a center frequency of 60$\,$MHz, in agreement with the frequency difference between the saturated absorption lines $^{85}$Rb F=3$\rightarrow$F'=(3,4) and $^{85}$Rb F=3$\rightarrow$F'=4.

Three aspects of our devices can be improved upon. First, a higher SHG efficiency can be achieved by either improving the fabrication technology and material quality or by using the materials with higher $\chi^{(2)}$ nonlineraity (such as LiNbO$_3$ \cite{wang2018ultrahigh,wolf2018quasi,luo2018semi}). Second, combining the SHG and electro-optic effect of AlN, it is possible to directly modulate the signal with integrated devices and simultaneously realize the Pound-Drever-Hall (PDH) locking schemes, further reducing the footprint of our experimental setup. Last, it is also possible to realize the SAS by packaging the PIC and the Rb atom cell \cite{Kitching2018}. For example, in Ref.~\cite{Hummon2018}, SAS has been demonstrated in a chip-packaged Rb cell.

In conclusion, we have studied the temperature and pump dependencies of SHG on an AlN photonic chip, and verified an approach to matching the working bandwidth of doubly-resonant SHG to a specific atomic transition. Taking the $^{85}$Rb D$_2$ transitions as an example, we demonstrated successful locking of an IR laser to the crossover line $^{85}$Rb F=3$\rightarrow$F'=(3,4) via an integrated frequency doubler. Our work provides proof of feasibility for integrated photonic platform utilization as a link between the frequency reference and the desired frequency.

\begin{acknowledgments}
National Key Research and Development Program of China (Grant No. 2016YFA0301300), National Natural Science Foundation of China (Grants No. 11874342, 11774110, and 91536219), and Anhui Initiative in Quantum Information Technologies
(AHY130000). J.S. and H.X.T acknowledge support from National Science Foundation (1640959) and Packard Foundation. This work was partially carried out at the USTC Center for Micro and Nanoscale Research and Fabrication.
\end{acknowledgments}

\end{document}